\documentclass[prl,twocolumn,10pt]{revtex4-1} 

\usepackage{amsfonts,amsmath,amssymb,graphicx}

\usepackage{color,times}

\definecolor{darkblue}{rgb}{0, 0, 0.8}
\usepackage[colorlinks=true, breaklinks=true, linkcolor=red, citecolor=darkblue, urlcolor=darkblue]{hyperref}

\graphicspath{{figures/}}

\renewcommand{\eqref}[1]{Eq.~(\ref{#1})}

\begin{document}

\title{Synthetic three-dimensional atomic structures assembled atom by atom}

\author{Daniel Barredo$^*$}
\author{Vincent Lienhard$^*$}
\author{Sylvain de L\'es\'eleuc$^*$}
\author{Thierry Lahaye}
\author{Antoine Browaeys}
\affiliation{Laboratoire Charles Fabry, Institut d'Optique Graduate School, CNRS, \\ 
Universit\'e Paris-Saclay, F-91127 Palaiseau Cedex, France}

\date{\today}

\begin{abstract}
We demonstrate the realization of large, fully loaded, arbitrarily-shaped three-dimensional arrays of single atoms. Using holographic methods and real-time, atom-by-atom, plane-by-plane assembly, we engineer atomic structures with up to 72 atoms separated by distances of a few micrometres. Our method allows for high average filling fractions and the unique possibility to obtain defect-free arrays with high repetition rates.  These results find immediate application for the quantum simulation of spin Hamiltonians using Rydberg atoms in state-of-the-art platforms, and are very promising for quantum-information processing with neutral atoms.
\end{abstract}

\maketitle

A major challenge of current research in quantum science and technology is to realize artificial systems of a large number of individually controlled quantum bits~\cite{Nielsen2000}, for applications in quantum computing and quantum simulation. Many experimental platforms are explored, from solid-state systems such as superconducting circuits~\cite{Devoret2013}, or quantum dots~\cite{Veldhorst2015}, to AMO systems such as photons, trapped ions, or neutral atoms~\cite{Kok2007,Buluta2011,Meschede2006,NatPhys2012, Gross2017}. The latter hold the promise of being scalable to hundreds of qubits or more~\cite{Weiss2017}. Quantum gas microscopes~\cite{Kuhr2016} allow for the realization of two-dimensional regular lattices of hundreds of atoms, and large, fully loaded arrays of $\sim$ 50 optical tweezers with individual control are already available in one \cite{Endres2016} and two dimensions~\cite{Barredo2016}. Ultimately, however, both for scaling to large numbers and for extending the range of models amenable to quantum simulation, accessing the third dimension while keeping single-atom control will be required. Here, we report on the realization of arbitrarily shaped three-dimensional arrays at unit filling, containing up to 72 atoms. These results open exciting prospects for quantum simulation with tens of qubits arbitrarily arranged in space, and show that realizing systems of hundreds of individually controlled qubits is within reach using current technology.

Large scale 3d quantum registers have been obtained using optical lattices with large spacings~\cite{Nelson2007}, which ease single site addressability and atom manipulation~\cite{Wang&Weiss2015}, but fully loaded configurations are not yet available. Here we use programmable holographic optical tweezers as an alternative approach to create 3d arrays of traps~\cite{Lee2016}. Holographic methods offer the advantage of a higher tunability of the lattice geometry, as the design of optical potential landscapes is reconfigurable and only limited by diffraction~\cite{Nogrette2014,Sturm2017}. 
\begin{figure}[h!]
\centering
\includegraphics[width=8.5cm]{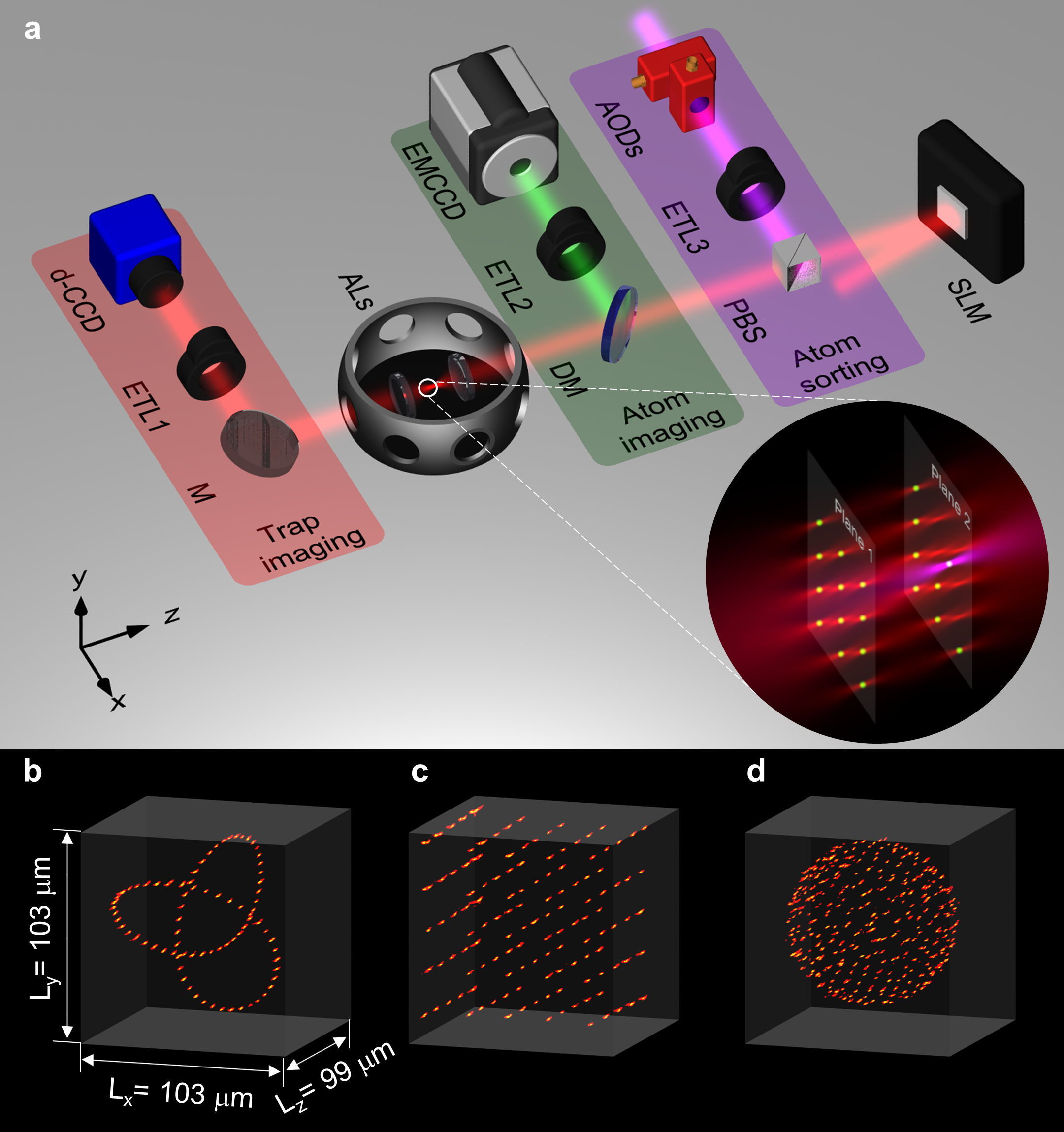}
\caption{\textbf{Experimental setup and trap images.} (a) We combine a SLM and a high-NA aspheric lens (AL) under vacuum to generate arbitrary 3d arrays of traps. The intensity distribution in the focal plane is measured with the aid of a second aspheric lens and a diagnostics CCD camera (d-CCD). The fluorescence of the atoms in the traps at 780 nm is separated from the dipole trap beam with a dichroic mirror (DM) and detected using an EMCCD camera. For atom assembly we use a movable tweezers (MT), superimposed on the trap beam with a polarizing beam-splitter (PBS). This extra beam is deflected in the plane perpendicular to the beam propagation with a 2d AOD and its focus can be displaced axially by changing the focal length of an electrically-tunable lens (ETL3).  The remaining electrically-tunable lenses (ETL1 and ETL2) in the camera paths allow for imaging different planes. The inset depicts the intensity distribution of the trap light forming a bilayer array (red) and the action of the MT on an individual atom (purple). (b-d) Intensity reconstructions of exemplary 3d patterns obtained from a collection of z-stack images taken with the diagnostics CCD camera. The regions of maximum intensity form a trefoil knot (b), a $5 \times 5 \times 5$ cubic array (c), and a C$_{320}$ fullerene-like structure (d). The dimensions of the images are indicated and are the same in all the examples.}
\label{fig:fig1}
\end{figure}
In our experiment~\cite{Nogrette2014}, arbitrarily designed arrays of up to $\sim 120$ traps are generated by imprinting a phase pattern on a dipole trap beam at 850 nm with a spatial light modulator (SLM) (Fig.~\ref{fig:fig1}). This phase mask is calculated using the 3d Gerchberg-Saxton algorithm simplified for the case of point traps~\cite{DiLeonardo2007}. The beam is then focused with a high numerical aperture (${\rm NA}=0.5$) aspheric lens under vacuum, creating individual optical tweezers with a measured $1/e^2$ radius of $\sim 1.1\, \mu$m and a Rayleigh length of $\sim 5 \, \mu$m. The intensity of the trapping light is measured, after recollimation with a second  aspheric lens, on a standard CCD camera. An electrically-tunable lens (ETL1)~\cite{footnote_optotune} in the imaging path allows us to acquire series of stack images along the optical axis $\hat{z}$ with which we reconstruct the full 3d intensity distribution. The imaging system covers a z-scan range of $200 \,\mu\rm{m}$.

Figure~\ref{fig:fig1} (b-d) shows some examples of patterns suitable for experiments with single atoms. The images are reconstructed using a maximum intensity projection method~\cite{Wallis1989} from 200 z-images obtained with the diagnostics CCD camera. With $\sim 3.5$~mW of power per trap we reach depths of $U_0/k_B \simeq 1\,{\rm mK}$, and radial (longitudinal) trapping frequencies of around 100~kHz (20~kHz). We produce highly uniform microtrap potentials (with peak intensities differing by less than $5\%$ rms) via a closed-loop optimization~\cite{Nogrette2014}.  $^{87}{\rm Rb}$ atoms are then loaded in the traps from a magneto-optical trap (MOT), with a final temperature of $25\,\mu$K. We detect the occupancy of each trap by collecting the fluorescence of the atoms at 780 nm with an EMCCD camera for 50~ms. A second tunable lens (ETL2) in the imaging path is used to focus the fluorescence of different atom planes.

\begin{figure}[t]
\centering
\includegraphics[width=8.5cm]{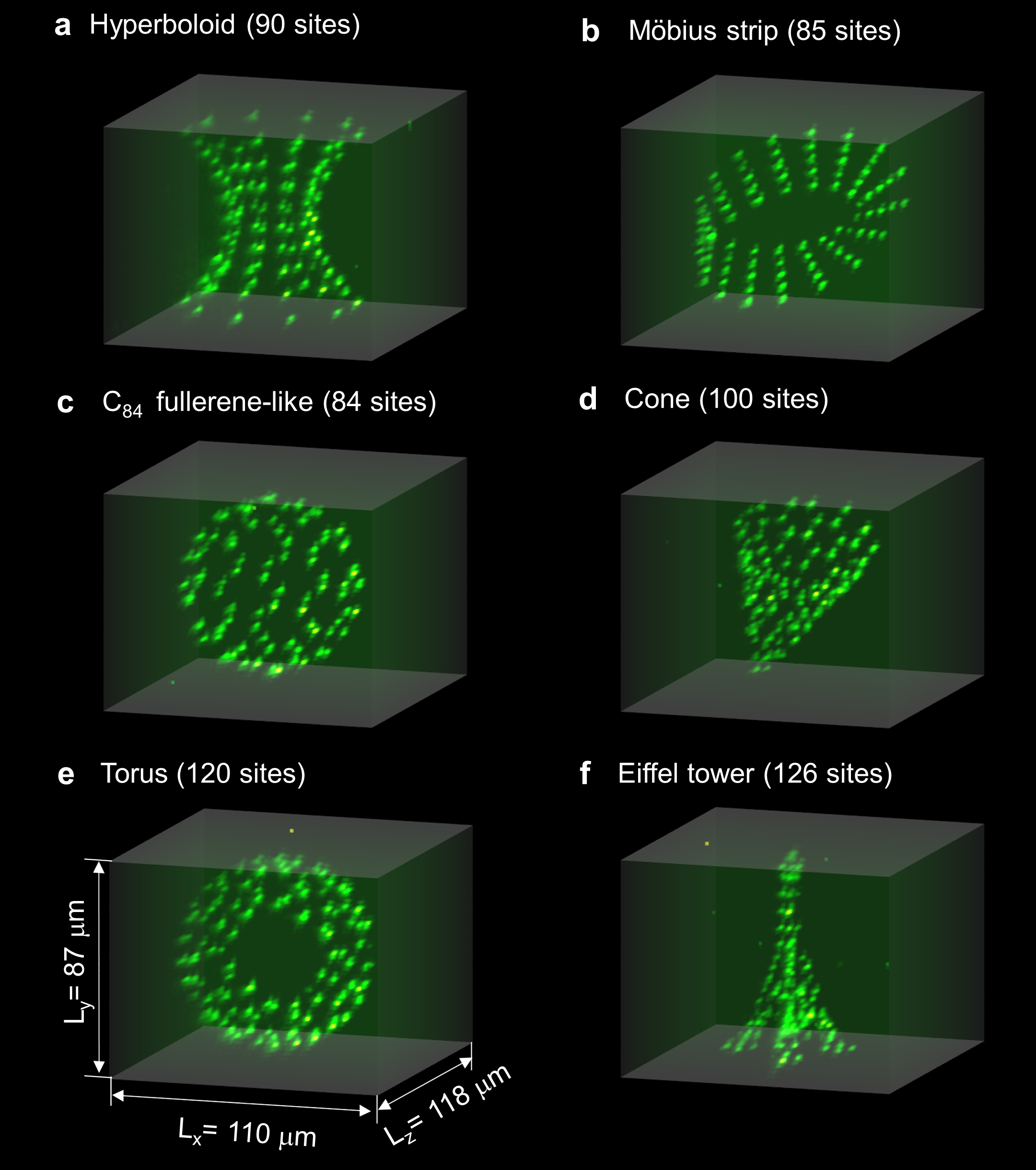}
\caption{\textbf{Single atom fluorescence in 3d arrays.} (a-f) Maximum intensity projection reconstruction of the average fluorescence of single atoms stochastically loaded into exemplary arrays of traps. The x,y,z scan range of the fluorescence is indicated and is the same for all the 3d reconstructions.}
\label{fig:fig2}
\end{figure}

In Fig.~\ref{fig:fig2} we show the fluorescence of single atoms trapped in various complex 3d structures, some of which are relevant to the study of non-trivial properties of Chern insulators~\cite{Beugeling2014,Ruegg2013,Ningyuan2015}. Each example is reconstructed from a series of 100 z-stack images covering an axial range of $\sim 120 \, \mu\rm{m}$. With no further action, these arrays are randomly loaded with a filling fraction of $\sim 0.5$; we thus average the fluorescence signal over 300 frames to reveal the  geometry of the structures. 

For deterministic atom loading, we extend our 2d atom-by-atom assembler~\cite{Barredo2016} to 3d geometries. For that, we superimpose a second 850-nm laser beam (with $1/e^2$ radius $ \sim 1.3\; \mu$m) on the trapping beam, which can be steered in the x-y plane using a 2d acousto-optical deflector (AOD) and in z by changing the focal length of a third tunable lens (ETL3). Combined with a real-time control system, this moving tweezers (MT) can perform single atom transport with fidelities exceeding 0.993~\cite{Barredo2016}, and produce fully loaded arrays by using independent and sequential rearrangement of the atoms for each of the $n_p$ planes in the 3d structures. 
\begin{figure*}
\centering
\includegraphics[width=18cm]{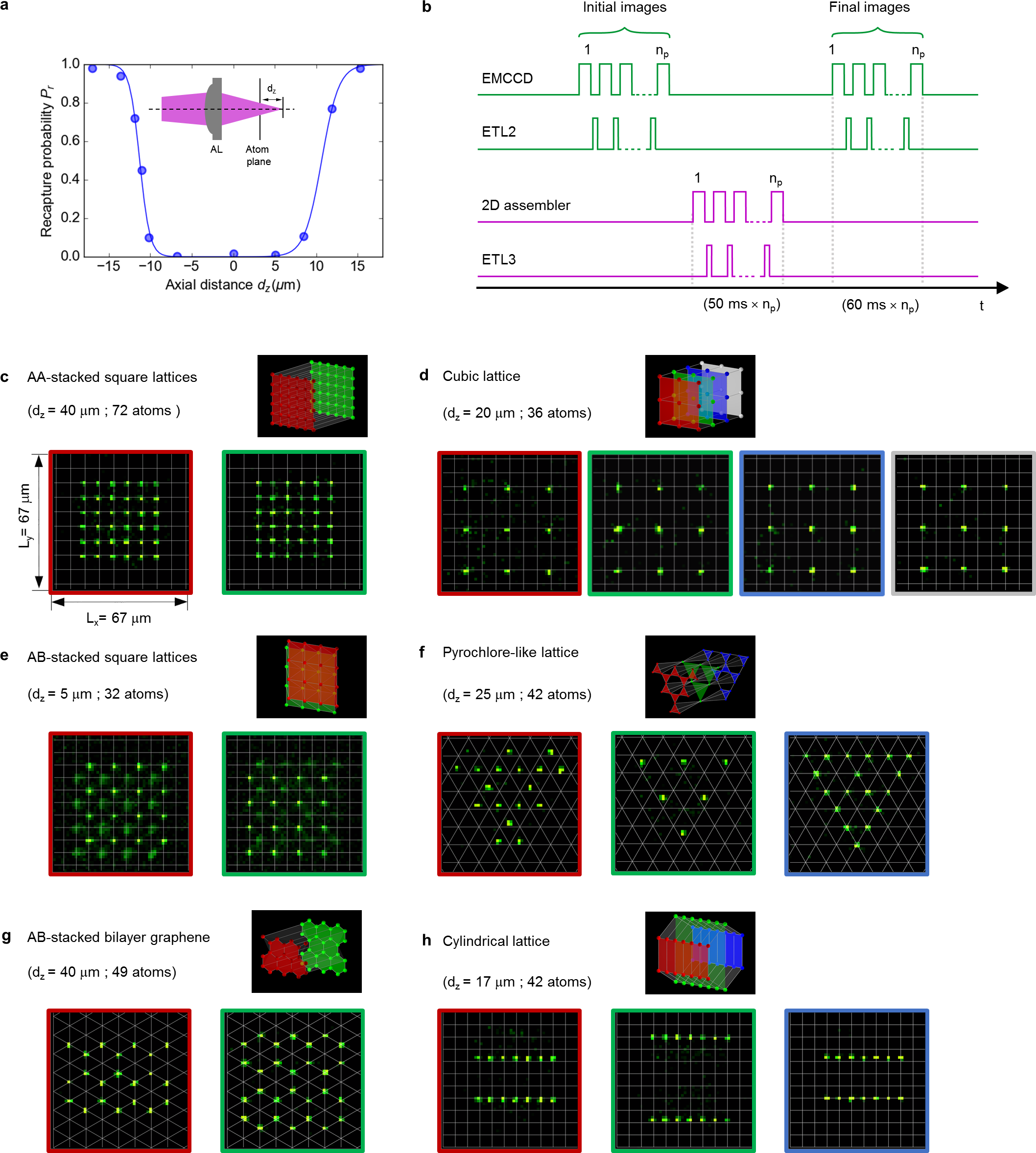}
\caption{\textbf{Fully loaded 3d arrays of single atoms.} (a) Recapture probability as a function of the axial distance between the MT focus and the plane of the atoms, for an experiment where we try to remove all the atoms from a 46-trap array. Error bars are smaller than the symbols size. The line is a guide to the eye. (b) Time control sequence of the experiment. We start the experiment by recording, sequentially, an image for each target plane. The analysis of the resulting $n_p$ images reveals the initial position of the atoms in the traps. The 2d atom assembler, in combination with an electrically-tunable lens (ETL3), arranges the atoms plane by plane. Finally, a new set of sequential images is collected to capture the result of the 3d assembly. (c-h) Gallery of fully loaded arrays with arbitrary geometries. All images are single shots. The models of the 3d configurations are shown for clarity; the various colors of the frame around the images encode successive atomic planes.}
\label{fig:fig3}
\end{figure*}

To explore the feasibility of plane-by-plane atom assembly, we first determine the minimal separation between layers so that each target plane can be reordered without affecting the others. To quantify this, we perform the following experiment in a 2d array containing 46 traps. We randomly load the array with single atoms and demand the atom assembler to remove all the atoms. We average over $\sim$ 50 realizations and then repeat the experiment for different axial separations between the MT position and the trap plane. The result is shown in Fig.~\ref{fig:fig3}a, where we see that for separations beyond $\sim 17\, \mu\rm{m}$ the effect of the moving tweezers on the atoms is negligible. This distance can be further reduced to $\sim 14\, \mu\rm{m}$ by operating the moving tweezers with less power, without any degradation in the performance of the sorting process. In a complementary experiment, where we fully assembled small arrays, we also checked that the assembling efficiency is not affected by slight changes (below $\sim 3\,\mu\rm{m}$) in the exact axial position of the MT.

We now demonstrate full loading of arbitrary 3d lattices using plane-by-plane assembly. We start by creating a 3d trap array which can be decomposed in several planes normal to $z$. In each plane we generate approximately twice the number of traps we need to load, such that we easily load enough atoms to assemble the target structure.  The sequence to create fully loaded patterns (see  Fig.~\ref{fig:fig3}b) starts by loading the MOT and monitoring the atoms entering and leaving the traps by sequentially taking a picture for each plane. We trigger the assembler as soon as there are, in each plane, enough atoms to fully assemble it. We then freeze the loading by dispersing the MOT cloud, and record the initial positions of the atoms by another series of z-stack images. The analysis of the images reveals which traps are filled with single atoms. We use this information to compute, in about 1 ms, the moves needed to create the fully loaded target array, and perform plane-by-plane assembling, changing the z-position of the MT. Finally, we detect the final 3d configuration with another series of z-stack images. 

Figure~\ref{fig:fig3}(c-h) shows a gallery of fully-loaded 3d atomic arrays, arbitrarily arranged in space. We can create fully loaded 3d architectures with up to 72 atoms distributed in several layers with different degrees of complexity. The selected structures include simple cubic lattices (d), bilayers with a square or graphene-like~\cite{Castro-Neto2009} arrangements (c,e,g), lattices with inherent geometrical frustration such as pyrochlore (f)~\cite{Bramwell2001}, or lattices with cylindrical symmetry (h), suitable e.g. to study quantum Hall physics with neutral atoms~\cite{Lacki2016}. The arrays are not restricted to periodic arrangements, and the positions of the atoms can be controlled with high accuracy ($<1 \,\mu\rm{m}$). The minimum interlayer separation we can achieve depends on the type of underlying geometry. This is illustrated in Fig.~\ref{fig:fig3}e, which shows the full 3d assembly of a bilayer square lattice (with a layer separation $d_z = 5\, \mu\rm{m}$). There, sites corresponding to the second layer are displaced by half the lattice spacing. Since traps belonging to neighbouring layers do not have the same x-y coordinates, there is no limitation for the minimum interlayer distance that we can produce. In both images we can observe a defocused fluorescence at inter-site positions due to atoms trapped in the neighbouring layer. In contrast, whenever traps are aligned along the z-axis (e.g. in Fig.~\ref{fig:fig3}d), we set a minimum axial separation of $\sim 17\,\mu\rm{m}$ to avoid any disturbance of the moving tweezers on the atoms while assembling neighbouring planes. However, for some trapping geometries, this constraint can be overcome by applying a small global rotation of the 3d trap pattern around the x or y axis, so that neighbouring traps do not share the same x,y coordinates. The range of interatomic distances we can achieve ($3-40\,\mu\rm{m}$) are well matched to implement fast qubit gates~\cite{Saffman2010}, or to simulate excitation transport~\cite{Browaeys2016} and quantum magnetism with Rydberg atoms, since interaction energies between Rydberg states at those distances are typically on the MHz range. The minimum interlayer spacing ultimately depends on the Rayleigh range of our trapping beam ($\sim 5\, \mu\rm{m}$), and could be further reduced, e.g. by using a higher NA aspheric lens.

Besides the unique tunability of the geometries it provides, our atom assembling procedure is highly efficient: we reach filling fractions of typically $0.95$. This measured efficiency is slightly dependent on the number of planes, and is mainly limited by the lifetime of the atoms in the traps ($\sim 10$~s) and the duration of the sequence (we typically need 60 ms per plane to acquire the fluorescence images and $\sim$ 50 ms per plane to perform atom sorting). The repetition rate of the experiment is about 1 Hz. The number of traps and the filling fraction of the arrays could be further increased with current technology: (i) the volume of the trap array and the maximum number of traps can be extended by increasing the field of view of the aspheric lens and the laser power; (ii) the lifetime of the atoms in the traps can realistically be increased by an order of magnitude; (iii) the repetition rate of the experiment can be increased by optimizing the atom assembler~\cite{Barredo2016}, in particular by transferring atoms also between different planes~\cite{Lee2016}; (iv) the initial filling fraction of the arrays could reach values exceeding 0.8 using tailored light-assisted collisions~\cite{Grunzweig2010,Lester2015}. Therefore, the generation of three-dimensional structures containing several hundred atoms at unit filling seems within reach, opening a wealth of fascinating new possibilities in quantum information processing and quantum simulation with neutral atoms.

\begin{acknowledgments}
We thank Andreas L\"auchli for useful discussions. This work benefited from financial support by the EU [H2020 FET-PROACT Project RySQ], by the `PALM' Labex (projects QUANTICA and XYLOS), and by the R\'egion \^Ile-de-France in the framework of DIM Nano-K.
\end{acknowledgments}

$^{\ast}$ D.~B, V.~L., and S. de L. contributed equally to this work.

\bibliographystyle{apsrev4-1_custom} 

%

\end{document}